\documentclass[aps,prb,superscriptaddress,twocolumn]{revtex4}
\usepackage{graphicx}
\begin{document}
\title{A versatile simulation of slow moving vortex lattices}
\author{Michael Dreyer}
\email{dreyer@lps.umd.edu}
\author{Jonghee Lee}
\author{Hui Wang}
\affiliation{Lab.\ for Phys.\ Sci., 8050 Greenmead Drive, College Park, MD 20740.}
\affiliation{University of Maryland, College Park, MD 20740.}
\author{Barry Barker}
\affiliation{Lab.\ for Phys.\ Sci., 8050 Greenmead Drive, College Park, MD 20740.}
\date{\today}

\begin{abstract}
We designed and implemented our own versatile simulation software in order to understand the velocity changes and track patterns observed in slow moving vortex lattices. The data was obtained from time series of STM images on NbSe$_2$ in a magnetic field range of $250-750$~mT. The main thrust is to explore possible driving mechanism and to test the effect of various defect configuration on vortex velocity and tracks. The simulation uses the full vortex--vortex and vortex--defect interaction. It allows for periodic boundary conditions as well as a repulsive sample edge. In early versions the time intervals for recalculating the force were estimated for individual vortices. This, however, led to a positional 'noise' of unacceptable temperature of the vortex lattice and was hence replaced by a global time step control. We were able to produce similar track and velocity patterns as well as local lattice distortions near point defects as observed by STM. A detailed analysis is, however, beyond the scope of this paper.
\end{abstract}

\maketitle

\section{Introduction}

The dynamic behavior of vortices in type-II-superconductors has been extensively studied both in experiment\cite{pardo_bishop_evlm2, pardo_bishop_evlm, troyanovski_kes_evlm, Nature.360.51, SupSciTech.14.729, jl_vm} and simulation. The simulations usually fall in one of two categories: Ginzburg-Landau-Theory\cite{PhysRevLett.73.3580, PhysRevLett.76.831, PhysRevLett.77.3208, PhysRevB.54.15372, PhysRevB.57.3073, PhysRevB.57.13861, JPhysCM.10.7429} or a 2D-molecular dynamics model\cite{PhysRevB.56.6175, PhysicaC.361.107, PhysRevB.64.064505, IntJModPhys.19.455}. The former is used to determine details of the vortex configuration, usually for a small area of the sample, from first principles. The second is used to study the behavior of a large number of vortices. The work presented in this paper falls within the second category. Among the observed effects which are sought to be explained are the peak effect\cite{JLowTPhys.34.409} of the critical current in current driven vortex lattices as well as the dynamic behavior of the vortex lattice as imaged by several techniques such as magnetic decoration techniques\cite{SCondSciTech.21.023001}, Lorentz microscopy\cite{JLowTPhys.131.941} or scanning tunneling microscopy\cite{troyanovski_kes_evlm, jl_vm}.

The main purpose of the simulations described in this paper is to compare the results to the behavior of vortices observed in high resolution scanning tunneling microscopy images obtained in a slowly decaying magnetic field in the range of 0.25--0.75 T\cite{jl_vm}. The images and the data derived from them have an unprecedented time resolution compared to the average velocity of the vortices (pm/s). Therefore, the data allows a detailed comparison in terms of track patterns, velocity variations and changes in lattice constant. The simulation will be mostly used within these field and velocity regimes. In this paper we analyze the the general behavior of the simulation in terms of collective effects, influences of the boundary conditions and defect interactions as well as give possible scenarios for the velocity patterns observed\cite{jl_vm}.

\section{The Simulation}

The simulation calculates the interaction of an ensemble of vortices with a static landscape of defects in two dimensions. The full vortex-vortex interaction is taken into account. The equation of motion for the $i$th vortex is given by:
\begin{equation} \eta \vec{v}_i = \sum_{j \ne i}^{N_{\mathrm{V}}} \vec{F}_\mathrm{VV}\left(\left|\vec{x}_i - \vec{x}_j \right| \right) + \sum_{j}^{N_{\mathrm{P}}} \vec{F}_\mathrm{VD}\left(\left|\vec{x}_i - \vec{x}_{\mathrm{D},j} \right| \right) + \vec{F}_\mathrm{global}
\end{equation}
Here $\vec{F}_\mathrm{VV}$ and $\vec{F}_\mathrm{VD}$ are the vortex--vortex and vortex--defect interactions, respectively. $\vec{F}_\mathrm{global}$ is an optional global driving force. The damping of the vortex motion $\eta$ is set to 1 (critical damping).

The main challenge of the calculation lies in the strong, long-range vortex--vortex interaction, the large number of vortices combined with the desired slow motion of the vortex lattice. In a first attempt, the recalculation of the vortex--vortex force was controlled for the individual vortex. The values of $\vec{F}_{\mathrm{V}_i\mathrm{V}_j}$ were stored in a table. Each vortex was allowed to travel a maximum distance $d_\mathrm{max}$ or evolve for a maximum time $t_\mathrm{max}$ --- whichever happened first --- before recalculating $\vec{F}_{\mathrm{V}_i\mathrm{V}_j}$. At this time $\vec{F}_{\mathrm{V}_j\mathrm{V}_i}$ is also updated. If the difference to the previous value exceeded a threshold, all forces for vortex $j$ were recalculated as well. Unfortunately, this approach led to large enough accumulative errors to be noticeable in form of lattice vibrations (see next section). Thus, the algorithm was changed to use a single time control representing the minimum value of all vortices.

\begin{figure}%
\includegraphics[width=\columnwidth]{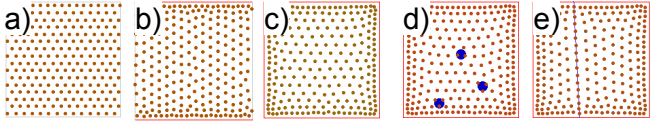}%
\caption{\label{Fig_setup}Final states of selected setup scenarios without external driving force. a) periodic boundary conditions, b) periodic in x, bound in y, c) repulsive boundaries, d) point defects, e) line defects.}%
\end{figure}

Two configurations were commonly used. The first utilizes periodic boundary conditions and a constant driving force. It was mainly used to study the interaction of vortices with point defects\cite{md_ld}. The size of the sample and the number of vortices control the orientation of the vortex lattice, especially for the vortex densities of interest. There are two basic possibilities for setting up a simulation under periodic boundary conditions. Either the number of vortices and the size of the sample are matched to form a perfect lattice or dislocations are introduced. In general, the number of vortices and the sample size were adjusted as to avoid lattice dislocations, while maintaining the target magnetic field. This adjustment leads to a horizontal alignment of the vortex lattice (cf Fig. \ref{Fig_setup}a). Furthermore, in the presence of defects, the direction of motion tends to align itself with a main axis of the lattice. The direction of motion is hence no longer identical to the direction of the driving force.

The second basic setup involved at least two repulsive boundaries (cf Fig. \ref{Fig_setup}b) and c). Here, a disordered layer of vortices forms at the edge of the sample making it very difficult to avoid lattice dislocations. The lattice is driven by removing or inserting vortices at a given point which is usually close to the edge of the sample and at random time intervals. This procedure generally introduces defects that move across the vortex lattice (see below). Also, for the field range considered in this paper, a significant motion along the sample edge is observed owing to the smoothness of the edge potential.   

\begin{table}
\begin{tabular}{|c|c|c|c|}
\hline
	Force   &Function   &$f_0$ [nN]& $d_0$ [nm]\\ \hline
	Vortex  &Bessel     & 30 -- 40 & 180 -- 200  \\ \hline
	Defect  &Gauss      & -40 -- 40& 20 -- 100   \\ \hline
	Boundary&Exponential& 4000     & 50        \\ \hline
\end{tabular}
  \caption{\label{Tab_params}Typical parameters used in the simulation.}
\end{table}
All forces are modeled as central forces using the form: $\vec{F}(\vec{d})=f_0\,\vec{n}_{\vec{d}}\,H(d/d_0)$. $H$ can be chosen to be a reciprocal, cubic reciprocal, exponential, Gaussian, or Bessel function. $\vec{d}$ is the distance between the two objects, $\vec{n}_{\vec{d}}$ its normalized direction. In the case of a line defect or a fixed boundary $d$ is the minimum distance. $f_0$ and $d_0$ are the force constant and decay length, respectively. The force and decay constant for the VV interaction are calculated from the magnetic field value based on theoretical curves of muon spin resonance measurements on NbSe$_2$\cite{sonier_brill_evcrad} following:
\begin{equation}f_{0,\mathrm{VV}}=\frac{\Phi_0^2\cdot t}{2\pi\mu_0\lambda^3}\label{eq:vv_force}\end{equation}
and
\[d_0=\lambda=\lambda_0\cdot\left(1+\beta\cdot\frac{H}{H_\mathrm{C2}(4.2\ \mathrm{K})} \right)\] with $\Phi_0=2.07\cdot 10^{-15}$ Wb, $\mu_0=1.26\cdot 10^{-6}$ Tm/A, $\lambda_0=144$ nm, $\beta=1.56$ and $H_\mathrm{C2}(4.2\ \mathrm{K})=2.13$ T\cite{sonier_brill_evcrad}. $t$ denotes the sample thickness (usually $t=500$ $\mu$m). Typical force function and parameters are summarized in table \ref{Tab_params}.

The setup of the simulation is controlled by an input parameter file which allows to easily change the parameters which include:
\begin{itemize}
	\item sample size, number of vortices
	\item vortex insertion/extraction rate
	\item periodic/fixed boundary condition
	\item fixed/time dependent driving force
	\item vortex--vortex interaction
	\item point/line defects
	\item control parameters ($d_\mathrm{max}$, $t_\mathrm{max}$,...)
\end{itemize}
Some examples of different simulation setups are displayed in Fig. \ref{Fig_setup}.

\section{Results}

\subsection{Collective lattice motion: Thermal noise}

\begin{figure}
	\includegraphics[angle=270,width=0.99\linewidth]{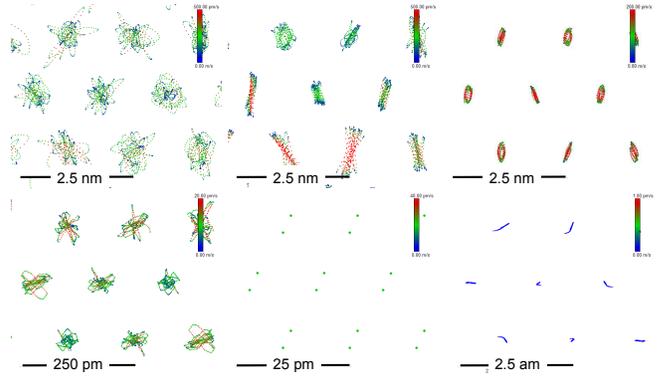}
\caption{\label{jitter}(Color online) Lattice vibrations caused by numerical errors in the simulation. The images show trajectories of neighboring vortices. The lateral scale bar refers to the excursions. The vortex--vortex distance remains $\sim 70$ nm. Top row: optimized algorithm for maximum evolution times of 20 s, 2 s, and 0.1 s respectively. Bottom row: common maximum evolution time step for all vortices (20 s, 2 s, and 0.1 s).}
\end{figure}
In early versions of the simulation attempts were made to optimize the calculation time. The highest computation cost lies in the calculation of the vortex--vortex interaction. By introducing an individual control of the evolution time before recalculating the forces on a given vortex the performance could be improved by a factor of up to $\sim 50$. However, a problem emerged in form of seemingly random lattice vibrations at amplitudes of $\delta p\sim 1$ nm. In a special set of simulations this motion was systematically studied (cf. Fig. \ref{jitter}). For large values of the maximum time step of $t_\mathrm{max}=20$ s, amplitudes of up to $A_\mathrm {Th}= 0.75$ nm and velocities of $v_\mathrm{Th}\le 500$ pm/s were observed. The values reduced to $A_\mathrm{Th}= 0.25$ nm and  $v_\mathrm{Th}\le 200$ pm/s for $t_\mathrm{max}=0.1$ s. The amplitude and velocities were unacceptable, so the code was changed to recalculate all forces at a common time interval. For large allowed time steps the problem prevailed although at smaller amplitudes ($A_\mathrm{Th}= 0.75$ pm, $v_\mathrm{Th}\le 40$ pm/s and $t_\mathrm{max}=20$ s) which could be reduced to $A_\mathrm{Th}< 1$ am and $v_\mathrm{Th}< 40$ fm/s at $t_\mathrm{max}=0.1$ s.

The reason for the motion were quasi random kicks to the vortex system, which may be due to rounding errors accumulated when calculating the sum of forces acting on a vortex. This sum has the particular challenge of adding numbers of vastly different magnitudes. Under these circumstances the order of the summation might lead to cases where $\vec{F}_{\mathrm{V}_1\mathrm{V}_2} \ne \vec{F}_{\mathrm{V}_2\mathrm{V}_1}$ which is likely responsible for the observed effect.  Fangohr et.\ al.\cite{JCompPhys.162.372} also showed, that under periodic boundary conditions the number of repetitions taken into account can have a significant effect on the outcome. This simulation originally used only the nearest neighbors either in the simulated area or in one of the repetitions. The range was later extended to the $2\mathrm{nd}$ nearest neighbor which led to a noticeable reduction in the noise level. The later, of cause, depends inversely on the size of the simulated area and vortex density. It becomes negligible for typical parameters used in the simulations described below.

The motion can be used to estimate an effective temperature of the system, although the calculations were performed at a nominal temperature of 0 K. Since the vortices are assumed to be massless the kinetic energy cannot be calculated. The variation of the potential energy $U$ from its minimum, however, can be expressed as an effective temperature:
\begin{equation}
T_\mathrm{eff}=\frac{1}{k_\mathrm{B}}\overline{\frac{\sum_{i=1}^N{\left(U(\vec{r}_i)-U(\vec{r}_0)\right)}}{N}}\label{eq:temp}\end{equation}
The potential for a single vortex is usually described by a sum over Bessel functions:
\begin{equation}
U(\vec{r})=-f_\mathrm{V}r_\mathrm{V}\sum_{j=1}^N K_0\left(\frac{|\vec{r}-\vec{r}_j|}{r_\mathrm{V}}\right)\label{eq:pot_sum}
\end{equation}

\begin{figure}[htb]
	\includegraphics[width=0.9\linewidth]{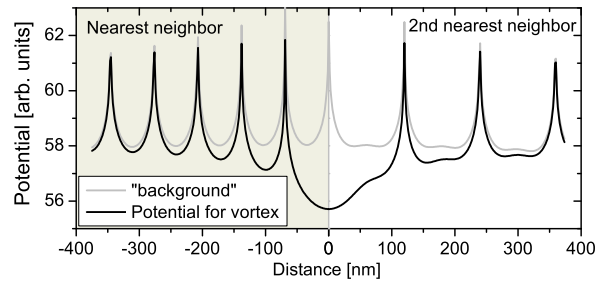}
	\caption{\label{pot_1}Potential for single vortex in the direction of the nearest and $2^\mathrm{nd}$ nearest neighbor at a magnetic field of 0.5 T.}
\end{figure}
\begin{figure}[htb]
	\includegraphics[width=0.75\linewidth]{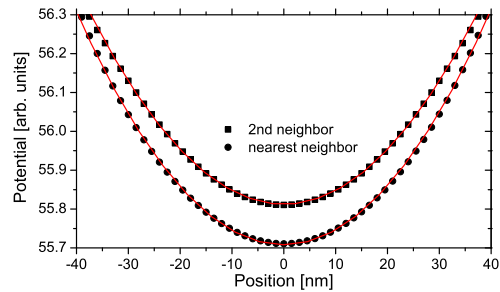}
	\caption{\label{pot_2}Parabolic fit of the potential near the vortex position. The plots are shifted vertically by 0.1 units for clarity.}
\end{figure}
The potential seen by a single vortex in the direction of the nearest and $2^\mathrm{nd}$ nearest neighbor are plotted in Fig. \ref{pot_1}. The potential around the vortex position is close to parabolic. Calculation of $\chi^2$ fits to a $2^\mathrm{nd}$ order polynomial (Fig. \ref{pot_2}, $y=a+bx+cx^2$) confirmed this assumption with correlation coefficients of $R\sim0.9997$. The coefficient $c\sim3.6\cdot 10^{14}$ dominates the $b\sim3\cdot 10^{4}$ term, and since we are interested in the order-of-magnitude of our temperature we can rewrite (\ref{eq:temp}):
\begin{equation}
T_\mathrm{eff}\sim \frac{f_\mathrm{V}r_\mathrm{V}c}{k_\mathrm{B}}\frac{\sum_{i=1}^N{\left|\vec{r}_i-\vec{r}_0\right|^2}}{N}\label{eq:temp2}\end{equation}
\begin{table}
\begin{tabular}{|c|c|c|}
\hline
	 Parameter&Nearest neighbor  &$2^\mathrm{nd}$ nearest neighbor\\ \hline
	a&55.7                &55.7              \\ \hline
	b&$-1.16\cdot10^{-10}$&$6.42\cdot10^{4}$ \\ \hline
	c&$ 3.52\cdot10^{14}$ &$3.77\cdot10^{14}$\\ \hline
  R&0.9998&0.9996\\
\hline
\end{tabular}
  \caption{\label{polpar}Parameters and correlation coefficient of the polynomial fit to the potential.}
\end{table}

\begin{figure}[htb]
	\includegraphics[width=0.75\linewidth]{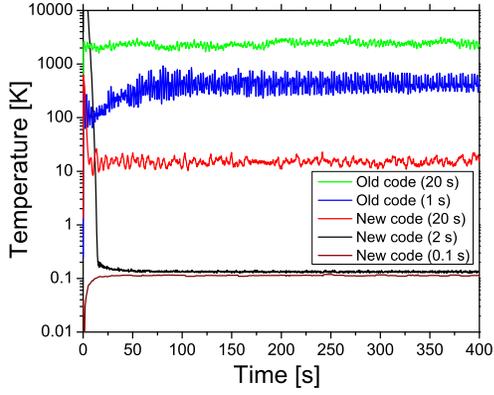}
	\caption{\label{temp}Temperature of the vortex lattice as calculated by equation \ref{eq:temp2}. The temperature shows a clear dependence on the algorithm and allowed $t_\mathrm{max}$}
\end{figure}

Fig. \ref{temp} shows the calculated temperatures of the simulations shown in Fig. \ref{jitter} (except for the second one) as a function of time. Starting with unrealistic values of several thousand K we finally achieved a temperature of $\sim 100$ mK. We used similar control parameters for subsequent simulations. Notably, the initial evolution of the temperature depends on the initial condition of the vortex lattice. This is most easily seen in the bottom two curves. For the top curve the initial positions of the vortices were chosen with a random offset $\le 10 \%$ of the lattice constant while the exact locations were used for the bottom curve. In the first case, the lattice cools down to the final temperature whereas it initially heats up in the second case. In either case it reaches thermal equilibrium after about 25 s.

\subsection{Point defects}

\begin{figure}[tb]
	\raisebox{0.35\linewidth}{\includegraphics[width=0.2\linewidth]{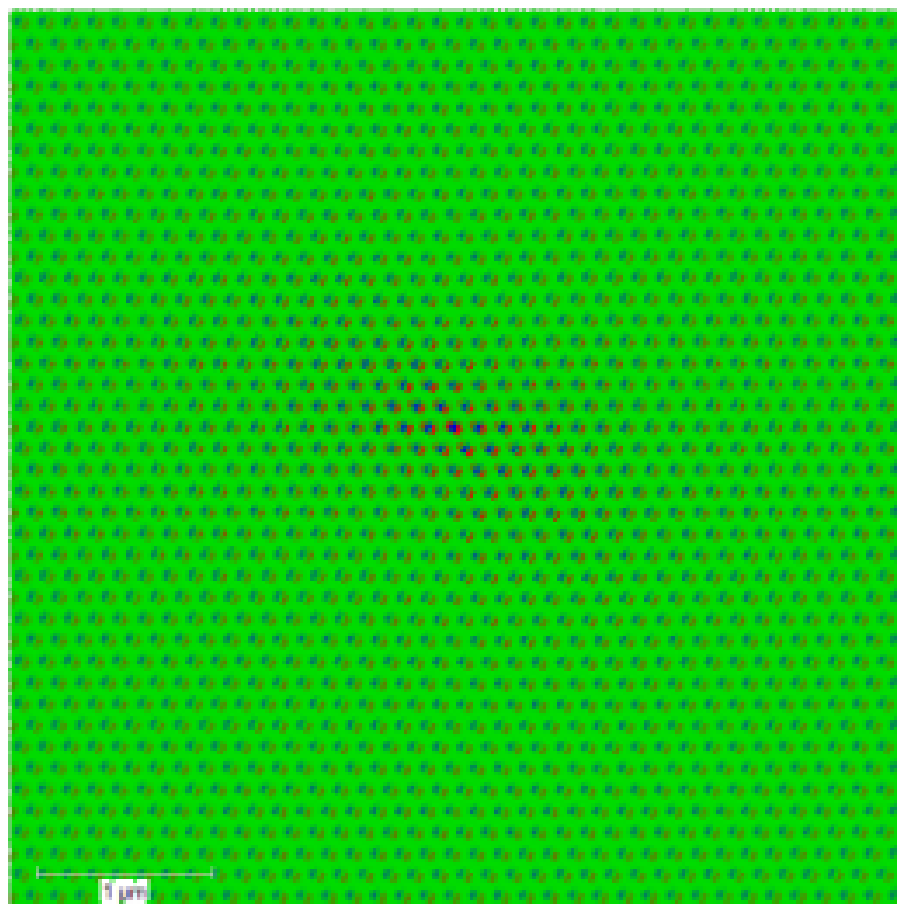}}\hspace{-0.2\linewidth}\raisebox{0.075\linewidth}{\includegraphics[width=0.2\linewidth]{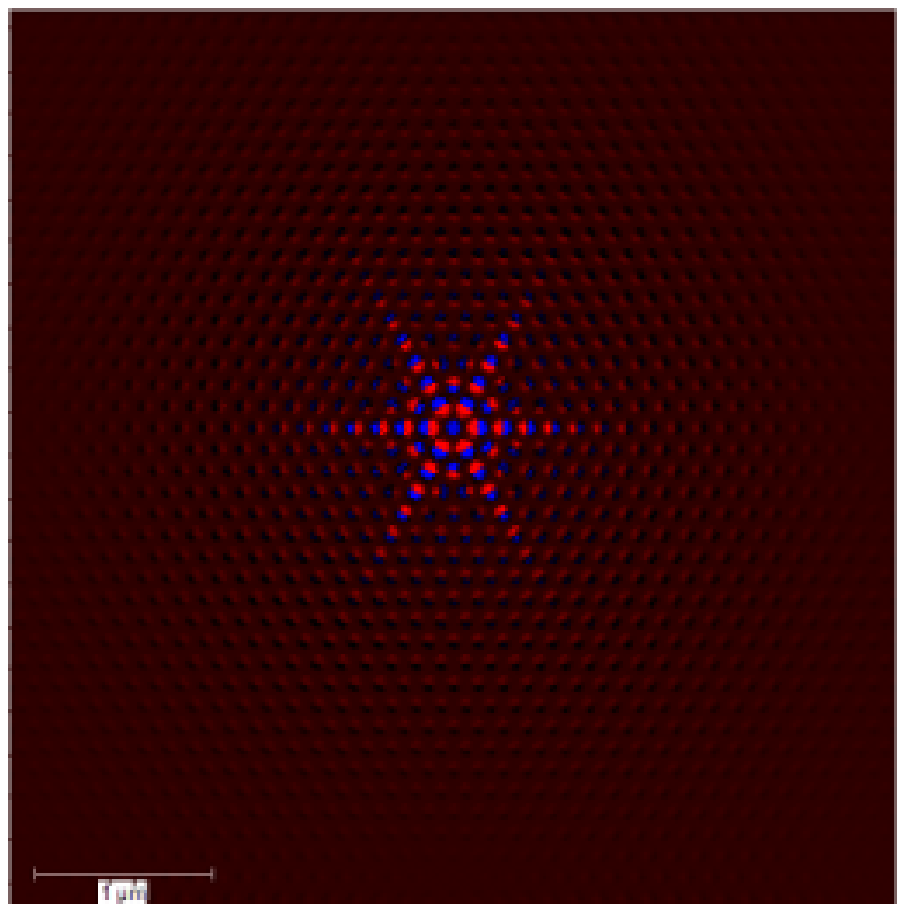}}\hfill\includegraphics[width=0.75\linewidth]{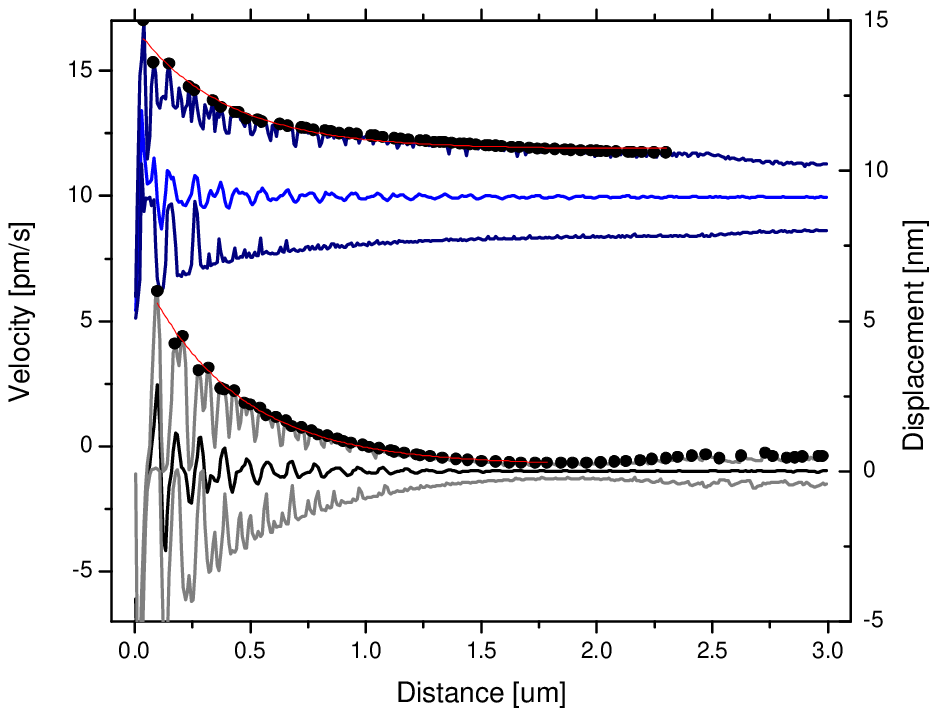}
	\caption{(Color online) Distance dependence of the velocity (top) and lattice distortion (bottom) for a point defect. The lines show the envelope and average curves. The dots represent local maxima and were used to estimate the decay of the variations as a function of distance by fitting an exponential decay.\label{pd_dist_dep}}
\end{figure}

We used single point defects to gain a basic understanding of the vortex--defect interaction. The goals were to ascertain the spacial extent of the velocity/distortion effects produced and to what degree an arrangement of one or more point defects could reproduce the observed velocity patterns. For the simulation described here, the number of vortices and the size of the area was adjusted to achieve a (except for the pinning center) strain and defect free vortex lattice, as described above. The density was chosen to reflect magnetic fields in the range of 60--500 mT. The driving force was aligned along one of the principal axis of the lattice with a magnitude of 10 pN, which would lead to a constant velocity of 10 pm/s without a defect.

\begin{figure}[tb]	 \includegraphics[width=0.45\linewidth]{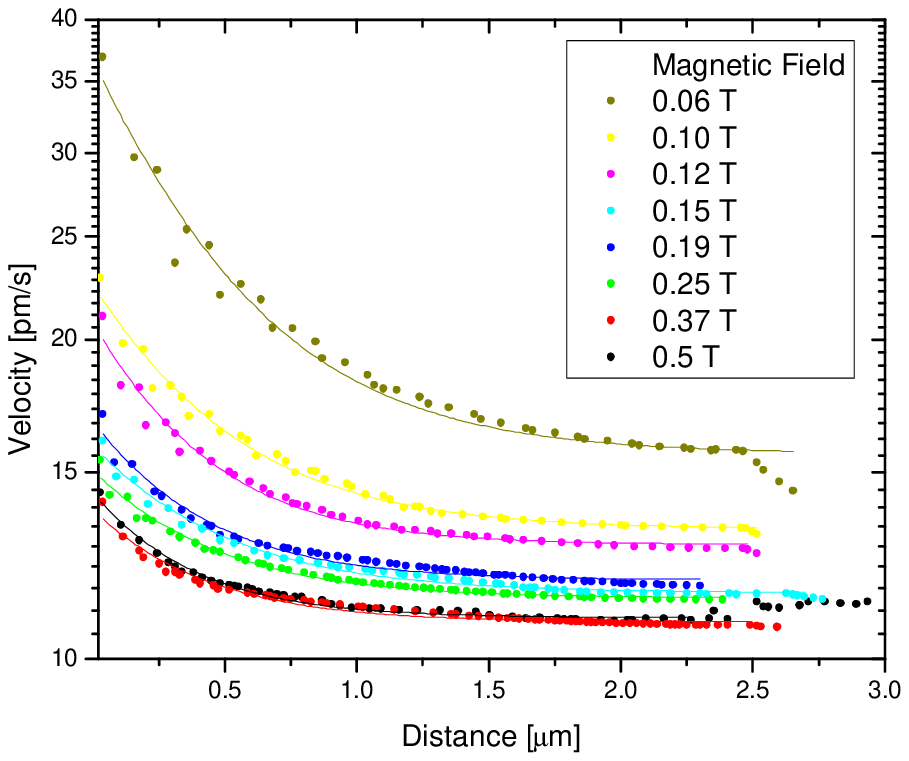}\includegraphics[width=0.45\linewidth]{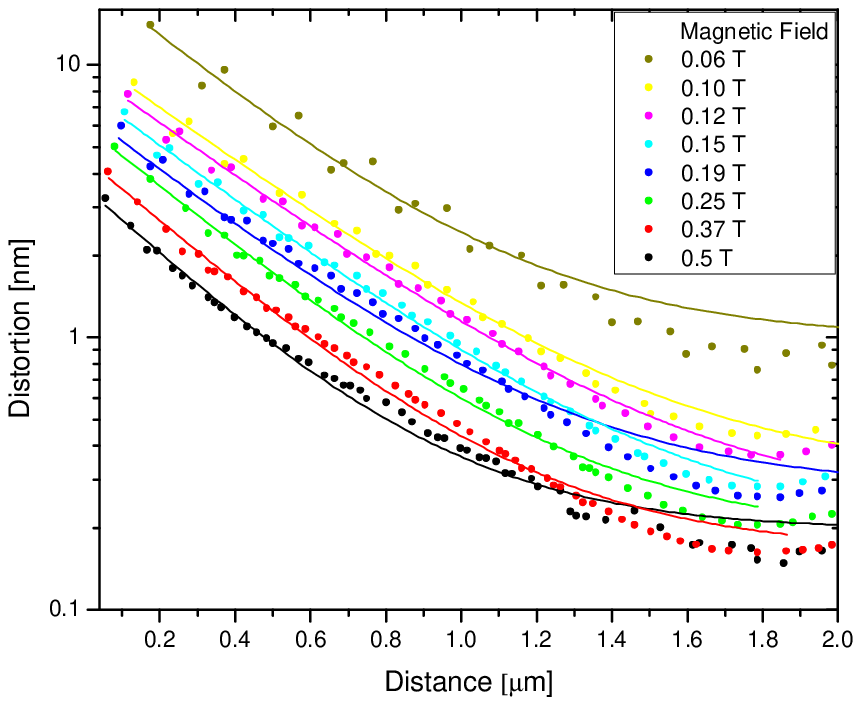}
	\caption{(Color online) Local maxima (cf. Fig. \ref{pd_dist_dep}) of velocity (left) and distortion (right) at different magnetic fields. The lines represent fits of an exponentially decaying function to the data.\label{pd_all_dist_dep}}
\end{figure}

\begin{figure}[tb]
	\includegraphics[width=0.75\linewidth]{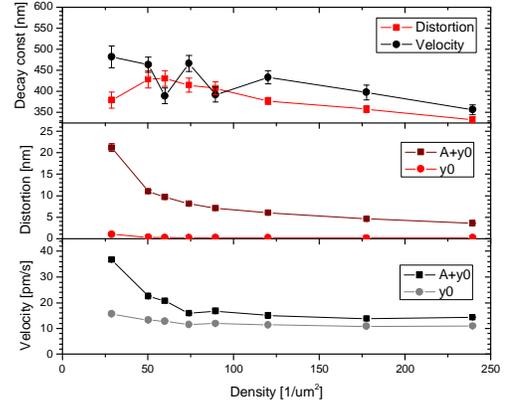}
	\caption{(Color online) Parameters of fits to an exponentially decay ($y=y_0+A_0\cdot\mathrm{e}^{-x/x_0}$) as a function of vortex density (magnetic field). The lines are a guide to the eye.\label{pd_par_vs_dens}}
\end{figure}

In the time domain we observed a periodic variation of the velocity in the range of 6 -- 14 pm/s --- largely due to the lattice periodicity combined with the periodic boundary condition. The distance dependence for a field of 125 mT is plotted in figure \ref{pd_dist_dep}. The graph shows the average and min/max values of the velocity and lattice distortion at a given distance from the defect. To calculate the width of the affected area we fitted the local maxima (black dots) to an exponential decay of the form $y=y_0+A_0\cdot e^{-x/x_0}$ using a least $\chi^2$ method. The local maxima and the fitted curves as a function of the magnetic field are depicted in figure \ref{pd_all_dist_dep}. The parameters as a function of vortex density are shown in figure \ref{pd_par_vs_dens}. The lattice shows a clear tendency to soften for lower magnetic fields as the amplitude of the velocity variation and the lattice distortion grow. This is not surprising since the force constant of the vortex--vortex interaction also diminishes with decreasing field according to (\ref{eq:vv_force}). The decay constant, however, shows only a weak dependence on field. The cause for the nonlinearity in $x_\mathrm{0,v}$ remains unclear at this point, as well as whether the declining trend of $x_\mathrm{0,d}$ continues for lower vortex densities. As noted elsewhere\cite{md_ld}, although single defects cannot explain the time dependent velocity patterns observed in the experiments, similar lattice distortions were observed which indicates the existence of point like pinning centers in real samples.

\subsection{Line defects}

\begin{figure}[tb]
  \begin{center}
	  \includegraphics[width=0.75\linewidth]{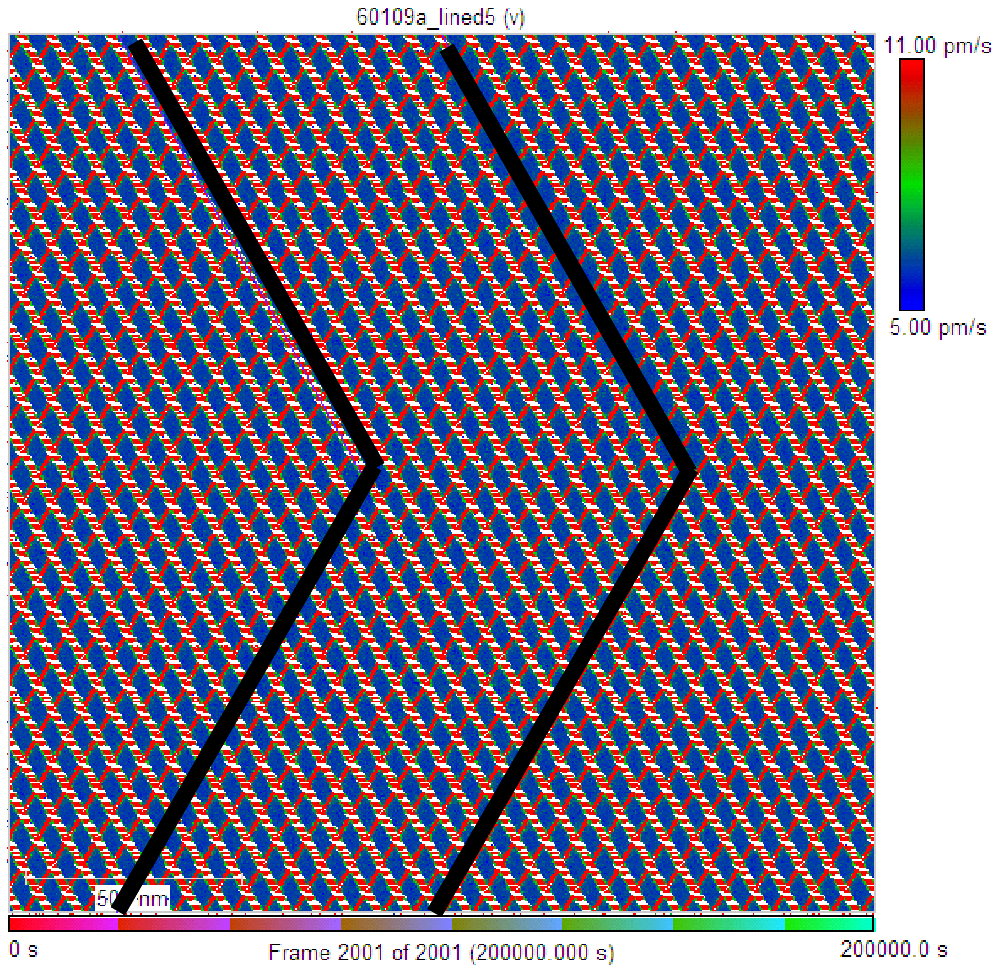}
	  \includegraphics[width=0.75\linewidth]{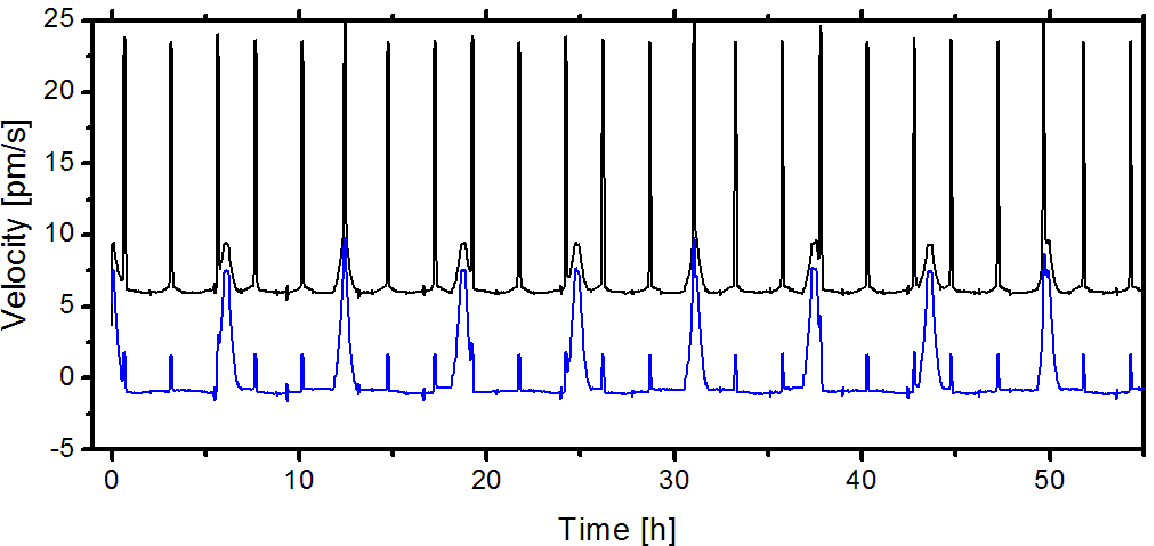}
	\end{center}
	\caption{(Color online) Velocity patterns caused by two periodic line defects. The image shows the emerging velocity patterns. The average velocity of all vortices parallel (top) and perpendicular (bottom) to the applied force is plotted against time.\label{fig_lines}}
\end{figure}
Line defects were introduced as a simple representation of extended defects in the host material such as step edges, dislocations or grain boundaries. The algorithm calculates the closest distance of a vortex to the line and the resulting force perpendicular to it. Since these types of defects are always present in a real material and are highly correlated, they present a possible explanation for the velocity patterns especially the spikes in the velocity we observed\cite{jl_vm}. However, this avenue was not pursued to a large extent. The setup using periodic boundary conditions also requires continuity of the lines at the edge and therefore restricts the orientation of the lines. Consequently, this leads to an overly strong periodic modulation of the average velocity (Fig. \ref{fig_lines}b).

In the case of repulsive boundaries and moderate force values, the line defects have a small influence on the overall pattern (see Fig. \ref{lat_disloc}, (bottom)). Only strong steps show a significant influence on the vortex motion over the effect of lattice dislocations (see below). For those cases, however, the step increasingly acts as another boundary, lining up the vortices and thereby introducing another slip plane in addition to the sample edge (cf. Fig. \ref{Fig_setup}e).

\subsection{Vortex lattice dislocations}

\begin{figure}[tb]%
\includegraphics[width=0.75\columnwidth]{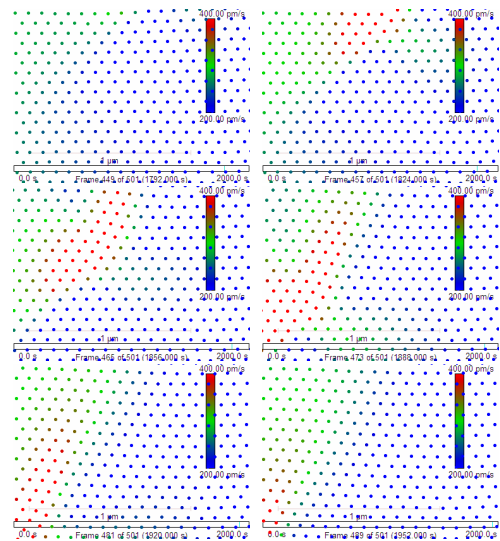}%
\caption{\label{lat_disloc_z}(Color online) Enlarged view of a lattice dislocation traveling through the field of view from the top right to the bottom left corner. The dislocation moves by transposing vortices along the direction of travel.}%
\end{figure}
\begin{figure}[tb]%
\includegraphics[width=\columnwidth]{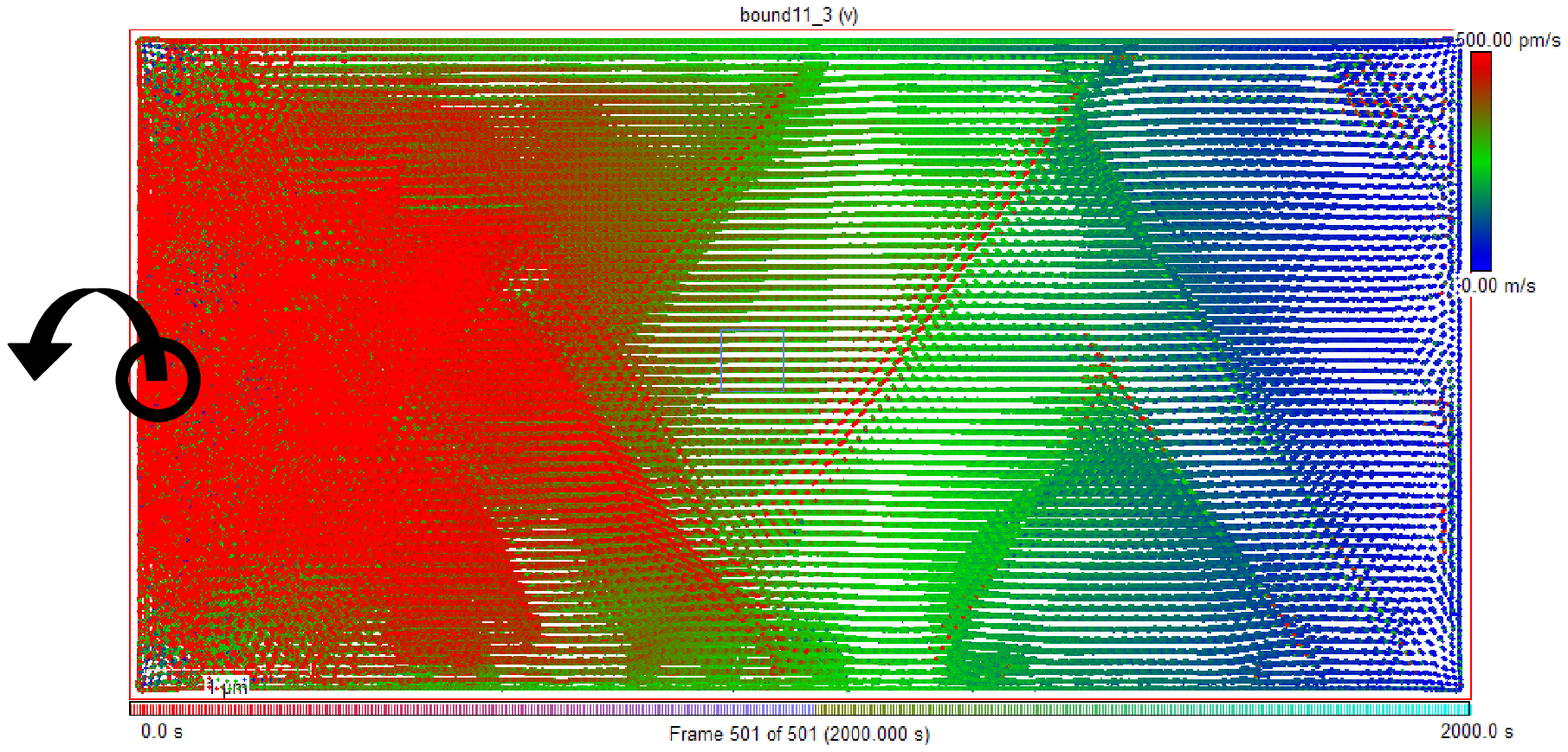}\\
\hfill\includegraphics[width=.93\columnwidth]{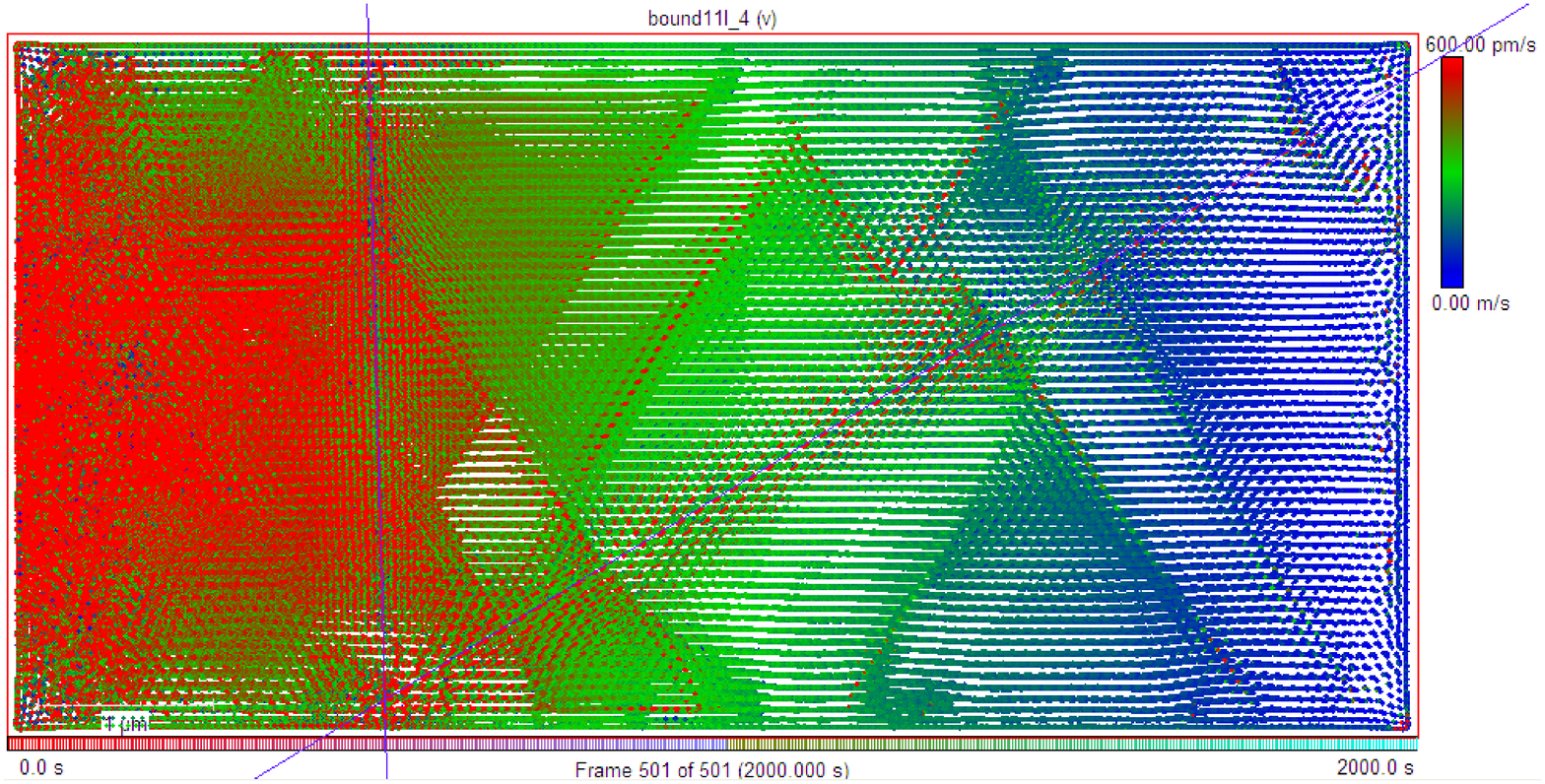}\\
\caption{\label{lat_disloc}(Color online) Lattice dislocation in large vortex array (7729 vortices) with repulsive boundaries. Vortices are extracted on the right side as indicated leading to a rearrangement of the lattice. The line like structures are vortex dislocation traveling through the lattice (see also Fig. \ref{lat_disloc_z}). The lower plot shows the result after adding two line defects. Even these correlated defects have little effect on the dislocation pattern.}%
\end{figure}
Dislocations within the vortex lattice proved to be an unexpected source of velocity variations. When removing vortices to drive the lattice, dislocations are generally created near the extraction site or the edge of the sample. These dislocation travel through the lattice at moderate velocities and lead to speed changes of the underlying vortex lattice along the way as shown in Fig. \ref{lat_disloc_z}. It should be stressed that the dislocations travels much farther and faster then the individual vortices do. In the simulation, the -- on a local level -- random motion of the dislocations lead to seemingly random velocity ``spikes'' of the vortex lattice. For large enough simulated areas the lattice should also break up into domains of local order. This could result in a plate tectonic of the domains (getting stuck/unstuck on neighboring domains) which in turn could lead to the velocity patterns we observed in our measurements. This proposed effect has not yet been observed in our simulations. Usually the number of vortices is too low to allow "`free floating"' domains which are not connected to the sample edge to develop.

Any defect introduced into one of these or similar large scale simulations has only a relatively small influence on the overall track patterns that emerge. This is shown in Fig. \ref{lat_disloc}: after the introduction of two line defects the track patterns remain somewhat similar. Due to the small area, the overall orientation of the vortex lattice is still governed by the boundary. For much larger areas (beyond our current computer capacity), this could easily change. 

\section{Conclusion}

In summary, we developed our own flexible 2D simulation to calculate the motion of an ensemble of vortices under various conditions in a landscape of defects. Based upon the average vortex displacement from its potential minimum we assert a typical temperature of the simulation of $T \sim 100$ mK. We explored the effect of a single point defect on the velocity patterns and lattice distortion and determined the radial extent of the influence. The amplitude increases with a decreasing magnetic field -- in accordance with a softening of the lattice -- whereas the decay constant only shows a weak dependence. Line defects as a model for surface steps or other correlated defects were also briefly discussed.

A new possibility for velocity variations within the vortex lattice was discovered in form of lattice dislocations traveling through the vortex lattice. For larger ensembles of vortices we could also envision locally ordered domains moving in unison producing velocity patterns by stick-slipping along their neighbors.

\bibliography{nbse2,vortex_sim,ourpapers,vortex_gen}
\bibliographystyle{apsrev}
\end{document}